\documentclass{article}
\usepackage[utf8]{inputenc}
\usepackage{authblk}
\usepackage{setspace}
\usepackage[margin=1.25in]{geometry}
\usepackage{graphicx,multicol}
\graphicspath{ {./figures/} }
\usepackage{subcaption}
\usepackage{amsmath}
\usepackage{lineno}
\usepackage{hyperref}
\usepackage{graphicx,xcolor}
\usepackage{setspace}
\usepackage{tocloft,cancel}
\usepackage{comment,color,soul,float} 
\usepackage{tikz,}
\usetikzlibrary{matrix,decorations.pathreplacing, calc, positioning,fit}
\usepackage{hyperref}
\usepackage{lineno}
\usepackage{multirow}
\usepackage{amssymb,amsmath}
\usepackage{colortbl}

\title{\huge \textbf{Revealing hidden bioimaging information by isotropic depolarization filtering}}

\author[1*]{Mónica Canabal-Carbia} 
\author[1]{Irene Estévez} 
\author[2]{Emilio González-Arnay} 
\author[1]{Ivan Montes-Gonzalez} 
\author[3]{Jose J. Gil}
\author[1]{Anrau Barrera}
\author[4]{Enrique García-Caurel}
\author[4]{Razvigor Ossikovski}
\author[5,6]{Ignacio Moreno}
\author[1]{Juan Campos}
\author[1]{Angel Lizana}
\begin{small}
\affil[1]{Grup d'Òptica, Dept. de Física, Universitat Autònoma de Barcelona, 08193, Bellaterra, Spain}
\affil[2]{Servicio de Anatomía Humana, Dept. de Ciencias Médicas Básicas, Universidad de la Laguna, 38200, Santa Cruz de Tenerife, Spain}
\affil[3]{Universidad de Zaragoza, Pedro Cerbuna 12, 50009, Zaragoza, Spain}
\affil[4]{LPICM, CNRS, Ecole Polytechnique, Institut Politechnique de Paris, 91120, Palaiseau, France}
\affil[5]{Inst. de Bioingeniería, Universidad Miguel Hernandez de Elche, 03202, Elche, Spain}
\affil[6]{Dept. de Ciencia de Materiales, Óptica y Tecnología Electrónica, Universidad Miguel Hernandez, 03202, Elche, Spain}

\affil[*]{Address correspondence to: monica.canabal@uab.cat}
\end{small}
\date{}
\begin{document}
\maketitle
\begin{center}
\Large\textbf{Abstract}
\end{center}

\normalsize We propose an imaging method to enhance and reveal structures within samples by using a polarization-based filter. This filter removes the isotropic content while amplifying the anisotropic component of depolarization. Whereas isotropic depolarization leads to a complete loss of polarimetric information, the anisotropic one is connected with intrinsic characteristics of samples. The filter has the capability to diminish the isotropic depolarization of samples, revealing their inherent information. 
As representative cases, we analyze the effect of the filter in heart and brain sections of animal origin. Results highlight the outstanding performance of the filter. In heart, myocardial and subendocardial structures are better visualized, whereas in the brain, fiber tracts are identified. These proves the significance of this filter in the medical field, paving the way to the early detection of pathologies. 
The methodologies here presented could be applied in a wide range of applications, providing a significant advance in polarization imaging where high isotropic depolarization response is present, this being a common scenario in nature.
\vspace{4mm}

\section{Introduction}
\label{sect:intro}

Polarimetric based methods are a powerful tool in a wide number of applications, as for instance in astronomy \cite{snik2013astronomical}, remote sensing \cite{tyo2006review}, environment \cite{kong2021polarization}, automated guided vehicles \cite{estevez2022urban}, botanical applications \cite{al2023characterization,rodriguez2022polarimetric}, biomedical applications \cite{ramella2020review,ivanov2021polarization},  among others. In the case of biomedical applications, polarimetric imaging and polarimetric based sample recognition are nowadays being used to study diverse human pathologies \cite{ivanov2021polarization,kupinski2018polarimetric,rodriguez2021polarimetric}. 
There are three main polarimetric properties that could be useful when studying samples: dichroism, retardance and depolarization. Depending on the nature and polarimetric features of the tissue to be inspected, one of these properties, or several, can store information about the sample after light-matter interactions. If the tissue presents a spatially heterogeneous polarimetric response, image contrast between structures can be highly improved through polarimetric methods \cite{canabal2023depolarizing,rodriguez2022automatic,gil2023polarimetric,van2018polarimetric,van2020depolarization}, this being of interest for imaging applications \cite{rodriguez2022automatic,gil2023polarimetric,van2018polarimetric,van2020depolarization}, as well as for tissues or pathologies automatic recognition \cite{ivanov2022polarization,majumdar2024machine}. 

Let us review the interest of these three main properties of materials in biomedical applications. Biological tissues are collagen rich structures, and collagen fibers are birefringent materials so retardance has arisen as an interesting tool to investigate different samples and pathologies, both through macroscopic or microscopic polarimetric techniques \cite{sun2014characterizing, sieryi2022optical,chue2018use,pardo2023wide, dong2017quantitatively,wang2016mueller}. In this sense, collagen density and fibrillar collagen orientation determine, respectively, the magnitude, orientation and alignment of  birefringence in biological tissues. Therefore, the amount, distribution, fiber orientation and alignment of fibrillar collagen are important factors underlying the properties of tissues, playing an important role in many diseases. For instance, the connection between collagen organization and birefringent properties has been used for the detection and progression study of different types of cancer \cite{clark2015modes,Keikhosravi:17}. Also nerve fibers and some proteins have an important birefringent response \cite{Sugiyama:15}. In addition, although the depolarization feature of samples was initially understood as a characteristic with little value in biomedical applications (it was regarded as noise to be minimized by some users), recent works have demonstrated that depolarizing channel encodes rich information of samples, as it is related with inherent features, as constituent units disorder, density, physical characteristics, etc. In this vein, due to the stronger scattering effects and organization changes on most tissues related to pathological processes, depolarization is a key method for the study and characterization of diverse human pathologies \cite{wood2010polarization,ivanov2021polarization,pierangelo2011ex,du2014mueller,kupinski2018polarimetric}. For instance, changes in cellular concentration  of tissues related to pre-cancer to cancer progression can be studied by means of depolarization variations\cite{du2014mueller}; also the variation of anisotropy levels and orientation disorder are directly related to heart pathologies such as infarction \cite{wood2010polarization}, cancer stages in different tissues as \textit{ex-vivo} human colon, skin, cervix and laryngeal cancer \cite{pierangelo2011ex,du2014mueller,kupinski2018polarimetric}. Therefore, anisotropic properties of samples can be related to different pathological stages. In this regard, the use of depolarizing metrics derived from the experimental Mueller matrix of the studied samples have demonstrated their interest in such applications. In turn, although observables related to samples dichroism, as diattenuation or polarizance \cite{chipman2018polarized}, are not commonly used in terms of biomedical samples imaging, due to the weak dichroic response of biological tissues \cite{ghosh2011tissue}, they have demonstrated to be helpful for tissues classification \cite{ivanov2022polarization,rodriguez2023optimizing} and, in the case of diattenuation this polarimetric feature has recently demonstrated its good performance in the study of brain tractography \cite{menzel2019diattenuation}. Moreover, they have proved a high interest for the study of plant samples, helping in the detection of chloroplasts and related organelles in plant species \cite{shtein2017stomatal}.

In this context, we recently published a study showing that depolarizing properties of samples can have two main origins: isotropic and anisotropic depolarization \cite{canabal2024connecting}. In this scenario, biological samples may exhibit anisotropic depolarization, isotropic depolarization or a mixture of both\cite{menzel2019diattenuation}.

In this work we show that in those samples where isotropic depolarization is a predominant effect, the anisotropic depolarization part may be present but hidden, this contribution being much more valuable in terms of image visualization and contrast enhancement. The discussion is conducted by considering some metrics suitable for the description of depolarizing samples, which have already demonstrated their interest regarding tissue imaging: the indices of polarimetric purity (IPP) \cite{van2018polarimetric,canabal2023depolarizing,rodriguez2022automatic,ivanov2021polarization}. Once this idea is set, we use this concept to implement an imaging filter with clear physical interpretation and very simple implementation. Such filter is based on removing the influence of one the IPP in the response of the sample. The interest of this new filter is tested on different biological samples, showing outstanding results in terms of sample visualization, and improving actual state-of-the-art. 

Finally, we want to note that the methods provided in this work are general and could be useful not only in biomedical applications, as motivated in this work, but in all those applications where polarization methods have already proved their interest in imaging or classification applications, as those stated at the beginning of this introduction. 

The present manuscript is organized as follows. In sec. \ref{sec:matb}, we present a brief summary of the isotropic and anisotropic depolarization concepts, the mathematical background related to the filter implementation in the Mueller matrix and the impact on some polarimetric observables after the isotropic filtering. Section \ref{sec:examples} provides the description of the two biological samples (transverse section of  an  \textit{ex-vivo} lamb heart and a coronal section across the frontal lobe of an \textit{ex-vivo} cattle brain sample) inspected in this work as well as their interest in the biomedical field (sec. \ref{sec:biologicalinterest}). The experimental results of the samples once applied the filter are shown in secs. \ref{sec:results_heart} and \ref{sec:results_brain}. To conclude, in sec. \ref{sec:conclusions} we provide the main conclusions of the work.
\section{Isotropic depolarization filter (IDF): Mathematical\\  
 Background}\label{sec:matb}
In this section, we present the mathematical background detailing a new tool for image processing based on polarization with high potential for vision applications, being specially suitable in samples showing depolarization, which is applicable to a large number of real scenarios. Specifically, we present a new concept for image polarimetric filtering which is applied to the experimental Mueller matrix images of samples, and exploits the fact that depolarizing samples may present isotropic and/or anisotropic depolarization.  

As discussed in  Ref. \cite{canabal2024connecting}, depolarization associated with a given uniformity in polarimetric properties is called anisotropic depolarization, while isotropic depolarization is caused by other effects that randomize the polarization state of light, such multiple scattering, but are unrelated to the polarimetric properties of the sample. 

In real samples, both isotropic and anisotropic depolarization usually occur simultaneously. The effect of isotropic depolarization is to reduce the contrast of images because it affects all elements of the Mueller matrix ($M$) in the same way. On the other hand, anisotropic depolarization affects the elements of $M$ differently and can therefore contribute to the improvement of the contrast in polarimetric images. Therefore, anisotropic depolarization seems to be more interesting for practical applications in polarimetric imaging than isotropic depolarization, and it is therefore interesting to find a way to separate their respective effects. In the following, we propose a method to filter out the isotropic content of the depolarization, which leads to a significant improvement in the visualization of sample structures, surpassing not only standard intensity images, but also current state-of-the-art polarimetric images.

To implement the polarimetric filter, we use the characteristic decomposition of $M$ \cite{gil2007polarimetric} in terms of the indices of polarimetric purity (IPP) observables \cite{san2011invariant}, because the IPP represent a suitable framework to separate isotropic and anisotropic depolarizing origins. Moreover, the characteristic decomposition has a clear physical interpretation that allows the filter to be implemented in an intuitive and simple way, which is useful in practical applications. In addition, this decomposition can also be useful for filtering polarimetric noise \cite{gil2016optimal}.

In the following we review the concept of isotropic and anisotropic depolarization in the context of characteristic decomposition (sec. \ref{subsec:charact}), and then we describe the filtering method and derive some observables of interest (sec. \ref{subsec:filter}).

\subsection{Isotropic and anisotropic depolarization content derived from the Characteristic Decomposition of the Mueller matrix} \label{subsec:charact}

The characteristic decomposition allows to write the $M$ of any depolarizer as the incoherent addition of different Mueller matrices. In fact, it separates $M$ as the contribution of four different matrices with physical interpretation (describing polarizing or depolarizing properties), and each one of these terms is weighted by one of the IPP or a linear combination of them. In particular, it can be expressed as follows \cite{gil2007polarimetric,san2011invariant}:
\begin{equation}
    M = P_1(m_{00}\hat{M}_{J0})+(P_2-P_1)(m_{00}\hat{M}_{1})+(P_3-P_2)(m_{00}\hat{M}_{2})+(1-P_3)(m_{00}\hat{M}_{3}),
\label{eq:characteristic_decomposition}
\end{equation}

\hspace{-6mm}where $m_{00}$ is an scalar value representing the mean intensity coefficient and the circumflex in the different matrices $M_{i}$ ($i=J0$, 1, 2 and 3) denotes the normalized matrix. 

Each one of the four matrices appearing in the incoherent addition in Eq. (\ref{eq:characteristic_decomposition}) has a particular physical meaning: $\hat{M}_{J0}$ represents the nondepolarizing features of $M$, $\hat{M}_1$ represents the portion of the medium that behaves as a 2D depolarizer, $\hat{M}_2$ represents the part of the medium that behaves as a 3D depolarizer and the term $\hat{M}_{3}$ gives the portion of the medium behaving as a perfect depolarizer \cite{gil2007polarimetric,gil2022polarized,gil2016structure}. For further interpretation, recall that a 2D depolarizer is a system that can be written as the incoherent addition of two specific pure Mueller matrices derived from $M$, whereas a 3D depolarizer is a system that can be written as the incoherent addition of three specific pure Mueller matrices derived from $M$, and a perfect depolarizer is a system that when a light beam interacts with it, regardless of its state of polarization, always transforms the incident polarization state into a fully unpolarized state (perfect depolarizers can be represented by the Mueller matrix $M_{\textit{perfect depolarizer}}=diag(1,0,0,0)$). 

Note that this additive scheme of interpretable elements give us valuable information about the polarimetric contributions leading to the final particular system represented by $M$.    
Importantly to our study, note that the significance (the weight) of each term in Eq. (\ref{eq:characteristic_decomposition}) is provided by a linear combination of the IPP parameters ($P_{1}$, $P_{2}$-$P_{1}$, $P_{3}$-$P_{2}$ and 1-$P_{3}$, respectively). Recall here that IPP are three polarimetric observables, $P_{1}$, $P_{2}$ and $P_{3}$, with values between 0 and 1, which are derived from the covariance matrix associated with a given $M$, and which provide quantitative information on the polarimetric randomness of the system \cite{san2011invariant,canabal2024connecting}. Therefore, the particular IPP combinations above-stated can be understood as metrics quantifying different polarimetric origins existing in a given sample. 

Stated the interest of IPP for an in-depth knowledge of depolarizing samples, in a previous work (Ref. \cite{canabal2024connecting}) we further studied the interpretation of IPP observables, by analyzing a collection of depolarizers consisting of the incoherent addition of easily interpretable devices (diattenuators and retarders), and inspecting the associated IPP values. We demonstrated that $P_{1}$ and $P_{2}$ were connected with anisotropic depolarization (originated by polarimetric anisotropy), and $P_{3}$ was connected with isotropic depolarization (perfect depolarizers). It was shown that depolarizing systems fully governed by anisotropic processes were characterized by $P_{3}=1$, regardless of the values of $P_{1}$ and $P_{2}$. Moreover, when $P_{3} < 1$ means that a given amount of isotropic depolarization was present. The limiting case of a  fully unpolarizing sample occurs when $P_{3}=0$. Readers interested in further evidence and discussion related to the connection of $P_{1}$ and $P_{2}$ with anisotropic depolarization, and $P_{3}$ with isotropic depolarization are addressed to Ref. \cite{canabal2024connecting}. 

In this framework, it is interesting to analyze how the characteristic decomposition changes when we consider systems with only anisotropic depolarization (i.e., with zero isotropic depolarization content) for which  $P_{3}=1$. Therefore, by imposing such condition in Eq. (\ref{eq:characteristic_decomposition}), the last term of the characteristic decomposition (corresponding to the perfect depolarizer contribution) becomes zero. Therefore, for this particular case Eq. (\ref{eq:characteristic_decomposition}) can be rewritten as \cite{canabal2024connecting}: 
\begin{equation}
   M_{P_3=1}=M_a = P_1(m_{00}\hat{M}_{J0})+(P_2-P_1)(m_{00}\hat{M}_{1})+(1-P_2)(m_{00}\hat{M}_{2}),
\label{eq:characteristic_decomposition_an}
\end{equation}

\hspace{-6mm}where $M_a$ denotes for depolarizing systems without isotropic depolarization. 

On the other hand, the last term of  Eq. (\ref{eq:characteristic_decomposition}) represents isotropic depolarization. In fact, by taking into account the weight of this last term in Eq. (\ref{eq:characteristic_decomposition}), i.e. $1-P_{3}$, it can be shown that the value of $P_3$ measures the proportion of anisotropic depolarization in a sample $M$ \cite{canabal2024connecting}. For instance, the condition $P_3=1$ is fulfilled  when no isotropic depolarization is presented in the sample and $P_3=0$ when  depolarization is fully isotropic. 
Moreover, due to the inequalities governing IPP, if $P_3=0$, the other IPP must be also zero, $P_1=P_2=0$ (see Eq. (S.4) of the Supplementary document). As a consequence, the characteristic decomposition of a system that only presents isotropic depolarization processes is written as \cite{canabal2024connecting}:
\begin{equation}
   M_{P_3=0}=M_{iso}=m_{00}\hat{M}_3,
\label{eq:characteristic_decomposition_is}
\end{equation}

\hspace{-6mm}where $M_{iso}$ denotes depolarizing systems fully governed by isotropic depolarization and represented by perfect depolarizers. Note that, in the equation describing isotropic depolarization processes the information is codify in only one term ($m_{00}\hat{M}_3$), whereas the anisotropic depolarization is described by three different terms (see Eq. (\ref{eq:characteristic_decomposition_an})).

A general system that may present both isotropic and anisotropic depolarization can be written as follows \cite{canabal2024connecting}: 
\begin{equation}
    M= P_3m_{00}\hat{M}_a+(1-P_3)(m_{00}\hat{M}_{iso}),
\label{eq:M_with_P3}
\end{equation}

\hspace{-6mm}where $M$ is divided in the anisotropic ($\hat{M}_A$) and isotropic ($\hat{M}_{iso}$) depolarizing contributions.

From Eq. (\ref{eq:M_with_P3}) we clearly realize that the $P_{3}$ metric associated to a Mueller matrix determines the portion of isotropic and anisotropic depolarization features in samples. Note also that, as a consequence of the inequality relation of IPP ($P_{1} \leq P_{2} \leq P_{3}$), if isotropic features of samples are predominant, $P_{3}$ takes small values, and thus, the valuable information of anisotropic depolarization, which is connected with physical properties of samples (through $P_{1}$ and $P_{2}$ channels and other related polarimetric observables \cite{gil2022polarized}) is mostly masked. This situation can be intuitively interpreted using the graphical representation of Purity Space \cite{gil2016components} (see sec. 3.1 of the Sup. doc.).

At this point, the following questions arise: (a) could it be possible to remove the influence of isotropic depolarization of a sample in order to highlight more physical properties of samples? and (b) if this is possible, would this be a powerful tool for image visualization of structures after removing isotropic depolarization? The main goal of this manuscript is to answer these questions and to prove, that: (a) it is very easy to isolate isotropic depolarizing features of a Mueller matrix (let us call it isotropic depolarization filtering) by using decomposition in Eq. (\ref{eq:M_with_P3}); and (b) the application of this filter paves the way for a new dimension of image processing based on polarimetric data. 
In the next subsection, the theoretical fundamentals of the proposed filter are described in detail. In turn, to highlight the potential of the method for practical applications, some experimental results and examples are provided in section \ref{sec:examples}.

\subsection{Isotropic depolarization filter} \label{subsec:filter}

The filter we propose consists in eliminating the isotropic ($(1-P_3)(m_{00}\hat{M}_3)$) term from $M$. The filtered Mueller matrix $M_a$ is implemented as follows: 

\begin{equation}
\begin{split}
    M_a = M -(1-P_3)(m_{00}\hat{M}_3) =  \\P_1(m_{00}\hat{M}_{J0})+(P_2-P_1)(m_{00}\hat{M}_{1})+(1-P_2)(m_{00}\hat{M}_{2}).
    \label{eq:m_filtrada}
\end{split}
\end{equation}

In this way, we isolate the anisotropic information in the new filtered matrix $M_a$, which corresponds to the anisotropic terms of the characteristic decomposition. In other words, we subtract the last term in Eq. (\ref{eq:characteristic_decomposition_an}) to the Mueller matrix $M$, to get $M_a$. 

Once the anisotropic information is isolated, we can study how this filtering affects the polarimetric observables that can be calculated from $M_a$ (and how they compare with the same observables derived from $M$). To do so, we can write the filtered matrix elements (those of $M_a$) in terms of the elements before the filtering. As the $\hat{M}_3$ matrix has the diagonal form $diag(1,0,0,0)$, it is straightforward to see that $m_{00}$ is the only element in $M$ affected by the filter: 
\begin{equation}
    m_{a00}=m_{00}-(1-P_3)m_{00}= P_3m_{00}.
    \label{eq:mic_filter}
\end{equation}

The rest of the elements of the filtered matrix (before normalization) are not affected by these operation, that is $m_{ai,j}=m_{i,j}$ for all elements except for $i=j=0$ (Eq. (\ref{eq:mic_filter})). To obtain the different polarimetric observables from $M$, it is necessary to normalize the matrix (divide every element of the matrix by $m_{00}P_3$). The normalized form of $M_a$ in terms of the elements before the filter, can be expressed as (the filtered polarimetric parameters are noted with the superindex '):
\begin{equation}
   M_a=m_{00}P_3
\begin{tikzpicture}[
  baseline,
  label distance=10pt 
]
\matrix [matrix of math nodes,left delimiter=(,right delimiter=),row sep=0.1cm,column sep=0.1cm] (m)  {
      1 & \frac{m_{01}}{m_{00}P_3}  &   \frac{m_{02}}{m_{00}P_3}  & \frac{m_{03}}{m_{00}P_3} & \\
       \frac{m_{10}}{m_{00}P_3} &  \frac{m_{11}}{m_{00}P_3}& \frac{m_{12}}{m_{00}P_3} & \frac{m_{13}}{m_{00}P_3} \\
       \frac{m_{20}}{m_{00}P_3} &  \frac{m_{21}}{m_{00}P_3} & \frac{m_{22}}{m_{00}P_3} & \frac{m_{23}}{m_{00}P_3} \\
      \frac{m_{30}}{m_{00}P_3} &  \frac{m_{31}}{m_{00}P_3} & \frac{m_{32}}{m_{00}P_3} & \frac{m_{33}}{m_{00}P_3} \\
       };
       
\draw[thin] (m-1-2.south west) rectangle (m-1-4.north east);
\draw[thin] (m-2-1.north east) rectangle (m-4-1.south west);

\node[
  fit=(m-1-2)(m-1-4),
  inner ysep=-2.5mm,
  label=above:$D'^{T}$
] {};
\node[
  fit=(m-2-1)(m-4-1),
  inner ysep=-2.5mm,
  label=below:$P'$
] {};

\end{tikzpicture}
\label{eq:relacion_matrices}
\end{equation}

From Eq. (\ref{eq:relacion_matrices}) we see that, since the $m_{00}$ element of the filtered matrix $M_a$ was affected by the IDF, the new normalized elements of the matrix are also affected. In particular, all the elements in the matrix have a dependency on $P_3$. The interest of filtering the isotropic part of the depolarization is particularly evident when it comes to image non-homogeneous samples which show a given spatial variation of their respective polarimetric properties across the imaged area. In this way, the value of $P_3$ will vary across the sample and therefore affect the observed value of the polarimetric properties differently. For instance, when we obtain the Mueller matrix image of a sample, we compute the value of each polarimetric characteristic pixel by pixel. Hereafter, we will explicitly write this point-to-point dependence of the polarimetric observables, $P_3(x,y)$.

Once the filtered matrix is obtained, we recalculated the polarimetric observables from this matrix in order to analyze the effect of subtracting the isotropic depolarization on them.  For a more detailed calculations see sec. 2 of the Supplementary Document.

\subsubsection{Non-depolarizing channels: Dichroic and Retardance properties}\label{nondep_channels}
In this subsection, we discuss the contrast improvement related to non-depolarizing observables (i.e., dichroic and retardance-based properties) by using the proposed $P_3$ based filter. Regarding to the sample characteristics related to dichroism, we mainly focus on diattenuation ($D$) and polarizance ($P$). These metrics can be obtained directly from the $M$ elements, as shown in Eq. (\ref{eq:relacion_matrices}). The relationship between the parameters before and after the filter can also be extracted  directly from Eq. (\ref{eq:relacion_matrices}). 
    \begin{equation}
    D'(x,y)=\frac{D(x,y)}{P_3(x,y)}, \quad P'(x,y)=\frac{P(x,y)}{P_3(x,y)}.
    \label{eq:filtered_cp}
\end{equation}

From Eq. (\ref{eq:filtered_cp}) we see that filtering $P$ and $D$ means dividing the original values of these observables by $P_3$. When the isotropic depolarization is high, the values of $P_3$ are low, and therefore filtering the isotropic component will imply a significant contrast enhancement in $P$ and $D$. In sec. \ref{sec:examples}  we show how this effect help not only to the contrast enhancement but also to reveal some structures hidden due to a high amount of isotropic depolarization.

The filtered retardance ($R'$) can not be obtained directly from the elements of $M_a$. To obtain the retardance observables from a Mueller matrix, we need to further processing the data, as for instance, by applying Mueller decompositions, as the Lu-Chipman, the Arrow or the Symmetric decompositions \cite{gil2022polarized}. The calculations from which those methods are based are not straightforward, so to obtain an analytical expression for the filter effect on $R'$ is not trivial and it is out of the scope of this manuscript. However, considering a heuristic approach based on the study of the filter effect in retardance images corresponding to a wide range of tested samples, we hypothesize that the contrast enhancement is notably lower than in observables related to dichroic and depolarization properties and that such filter effect can be considered negligible for the retardance case.

\subsubsection{Depolarizing channels}\label{dep_channels}
Different sets of depolarization related parameters can be deduced from $M$. It has been demonstrated that the IPP give the best performance and provide fundamental information about the origin of depolarization in biological tissues \cite{canabal2024connecting}. In addition to IPP, here we discuss the effect of the IDF on two more metrics related to depolarization: the depolarization index ($P_{\Delta}$), which is a global indicator of the depolarization produced by a sample and the degree of spherical purity ($P_s$), which measures the contribution to depolarization that is not directly related to the dichroic properties \cite{gil2022polarized,chipman2018polarized}.

We start by calculating the effect of the IDF on the IPP. The IPP are obtained as linear transformations of $M$ and can not be directly obtained from $M$ elements, as these indices are linear combinations of the covariance matrix, $H(M)$, eigenvalues $\lambda_{i}$. \cite{gil2022polarized,san2011invariant,cloude1986group}. Therefore, to obtain the filtered IPP we need to calculate the covariance matrix of $M_a$ in Eq. (\ref{eq:m_filtrada}), i.e. $H(M_a)$. Each element of the covariance matrix can be calculated as a linear combination of different elements of $M$, and importantly, the $m_{00}$ element is only present in the diagonal of $H$ (see Eq. (S.1) of the Sup. doc.) \cite{gil2022polarized}. Therefore, the diagonal elements of $H$ will be the only ones affected by the filter, and $H(M_a)$ can be obtained as:
\begin{equation}
    H(M_a) = H(M)-(1-P_3)m_{00}\mathbb{I},
    \label{eq:filtered_h}
\end{equation}

\hspace{-6mm}where $\mathbb{I}$ is the identity matrix. 

The next step is to obtain the eigenvalues of $H(M_a)$, $\lambda'_{i }$, to calculate the filtered IPP (i.e., $P'_{1}$, $P'_{2}$ and $P'_{3}$). To this aim, we can relate the covariance matrix before, $H(M)$, and after, $H(M_a)$, the filter by using Eq. (\ref{eq:filtered_h}) through the diagonalization calculation: 
\begin{equation}
\begin{split}
    H(M_a)-& \lambda'  \mathbb{I}=0 \rightarrow H(M-(1-P_3)m_{00})-\lambda \mathbb{I}=0 \\ 
    &H(M)-(\lambda' +(1-P_3)m_{00})\mathbb{I}=0,
\end{split}
\label{eq:relationlambdas}
\end{equation}
\begin{equation}
    \lambda=\lambda_a+(1-P_3)m_{00},
    \label{eq:rellamb}
\end{equation}

\hspace{-6mm}where $\lambda$ and $\lambda'$ correspond to the eigenvalues of $H(M)$ and $H(M_a)$ respectively. Eq.(\ref{eq:rellamb}) allows to calculate the relationship between IPP before and after the filtering. The IPP are defined as a function of the eigenvalues as follows \cite{san2011invariant,ossikovski2019eigenvalue}:
\begin{equation}
    P_n = \frac{1}{trH}\sum_{k=1}^nk\Delta \lambda_k, \quad n=1,2,3
\end{equation}

\hspace{-6mm}where $\Delta \lambda_k = \lambda_{k-1}-\lambda_k$ and $tr\mathbf{H}$ stands for the trace of the covariance matrix $H$ where $tr\mathbf{H}=m_{00}$. Therefore, regarding Eq. (\ref{eq:relationlambdas}), the relation between eigenvalues, $\lambda_i =\lambda'_i+P_3$ (i=0,1,2,3), and the traces $trH(M_a)=m'_{00}=P_3m_{00}$, we can write the filtered IPP as:
\begin{equation}
\begin{split}
    P'_n  = \frac{1}{trH(M_a)}\sum_{k=1}^nk\Delta \lambda_a = & \frac{1}{P_3trH}\sum_{k=1}^nk\Delta \lambda'_k= \frac{1}{P_3trH}\sum_{k=1}^nk\Delta \lambda_k = \frac{1}{P_3}P_n\\
   & P'_n (x,y) = \frac{P_n(x,y)}{P_3(x,y)},
\label{eq:relation_ipp}
\end{split}
\end{equation}

\hspace{-6mm}where $\Delta \lambda_a=(\lambda_{k-1}-P_3)-(\lambda_k-P_3)=\lambda_{k-1}-\lambda_k= \Delta \lambda_k$. Inspecting Eq. (\ref{eq:relation_ipp}) we can observe the same effect on the filtered variables than in Eq. (\ref{eq:filtered_cp}); the value of the filtered index increases by an amount given by $1/P_3(x,y)$, in the same way as the parameters presented in Eq. (\ref{eq:filtered_cp}).

In addition, we present the effect of the IDF on ($P_{\Delta}$) and ($P_s$). Both of these metrics can be obtained directly from the components of $M$ \cite{gil2022polarized} and, the effect of the filter on these parameters is the same as for the IPP and the dichroic parameters:
\begin{equation}
   P'_{\Delta}(x,y) = \frac{P_{\Delta}(x,y)}{P_3(x,y)},  \quad P'_s(x,y) = \frac{P_s(x,y)}{P_3(x,y)}.
   \label{eq:pdeltaps}
\end{equation}
\subsection{Discussion}

In Secs. \ref{nondep_channels} and \ref{dep_channels} we discussed how the visibility of images representing polarimetric observables \cite{kong2021polarization,al2023characterization,ramella2020review,estevez2022urban}, can be significantly improved using the proposed subtraction of the effect of isotropic depolarization. For most of these observables, their visibility is increased by a factor proportional to $1/P_3$ when the IDF is applied. Thus, this contrast enhancement is particularly relevant for small values of $P_3$, which is the typical case for polarimetric images of biological tissues.

Under this scenario, the potential of the IDF for applications in biophotonics is discussed in the following sec.\ref{sec:examples} by analyzing the contrast enhancement (and biological structure unveiling capability) obtained when applying our proposed IDF for studying diverse examples of biological tissues of animal origin. As additional information, if the readers are interested in a more visual interpretation of the filter, we address them to the Supplementary material (sec. 3), where we provide a physical interpretation of the filter in terms of depolarizing spaces \cite{gil2016components}.
\section{Filter application to the analysis of biological samples}\label{sec:examples}

As discussed in previous section, the contrast enhancement associated to the isotropic depolarization filter application on the study of a given sample, it is higher as lower the $P_3$ value is. In other words, the IDF is specially suitable for samples showing rich isotropic depolarization behaviour, which masks other polarimetric features of the sample. This scenario is common when dealing with biological samples, as for instance it is the case of soft tissues, usually presenting very low values for $P_3$ metric (see Table 1 in the Supplementary material). For this reason, in this section we present different examples of the IDF benefits when applied for the study of different soft tissues. 

In particular, in the following we present the application of the IDF for the study of two different tissues: sections of \textit{ex-vivo} heart and brain measured at 625 nm and 470 nm, respectively. In section \ref{sec:biologicalinterest}, we provide a brief description of the samples and the motivation of this choice due to their importance in the field of medicine. The Mueller matrices, $Ms$, corresponding to the stated samples were experimentally obtained by means of a complete Mueller matrix image polarimeter described in the sec. 4 of the Sup. doc.. In addition, for completeness, in the supplementary material we also provide the results obtained for other polarimetric observables and some examples of other samples, highlighting as well the potential of the filter for medical applications. 

\subsection{Biological samples: description and interest}\label{sec:biologicalinterest}

We have obtained excellent results in terms of image contrast enhancement after applying the filter to the polarimetric images of a number of soft tissue samples. For the sake of brevity, we will discuss two representative examples in this section: (1) heart and (2) brain tissue sections; both from \textit{ex-vivo} animal. We have chosen these two examples because of their great interest in the medical field and because, there is currently no standard non-invasive gold technique to characterize certain structural features and/or pathological tissues associated to these samples. 

In the case of the heart, cardiovascular disease is the number one cause of death in the world, leading to around the $15\%$ of total worlds deaths and has increased in more than 6 million cases in the last 20 years \cite{who}. The early diagnosis of structural cardiac abnormalities can help in the detection, prevention ad treatment of heart malfunctioning and unfavorable cardiovascular events such as infarction \cite{karamitsos2020myocardial}. Analyzing cardiac tissue remodeling at an early stage can be lifesaving. Currently, the gold standard technique for detecting myocardial fibrosis is the endomyocardial biopsy \cite{karamitsos2020myocardial}. This technique is an invasive method, and it is practically infeasible in daily clinical routine \cite{graham2017imaging}. In this sense, different studies have demonstrated the usefulness of polarimetry in the heart structural variations  and in differentiating between healthy, infarcted and regenerated myocardial tissue \cite{ahmad2017review,wood2010polarization}. Particularly, depolarization has arisen as an indicator of these structural changes, being directly correlated with the anisotropic structure of the heart tissue components \cite{ahmad2017review}. Interestingly, the approach here presented also allows to follow the trajectories of the subepicardial coronary vessels and generate contrast between their walls and their lumens. This feature may be suitable for the clinical context, as it would provide a tool for surgeons without the use of intra-surgical arteriography (see Fig. S6 of the Supplementary Material).   

In the case of the brain, the study of brain connectivity and its functional expression is probably the present frontier of applied neuroscience, as there is no gold-standard technique for pathway mapping apart from \textit{peri-mortem} tract-tracing injections that are not ethically suitable for the study of human connectivity \cite{charvet2023mapping}. Techniques such as histology or ultrastructure-based methods can distinguish fiber orientation but are not useful for tracing long range tracts \cite{agrawal2011josef}. It has been demonstrated the ability of Mueller matrix polarimetry  to classify large tracts that are known to have different orientations (internal capsule, cerebral peduncles, fimbria, medial lemniscus, optic tract) and to resolve the limits between grey and white matter \cite{gil2023polarimetric,menzel2019diattenuation}.

Recently, polarimetry has demonstrated its capabilities to study in an easy and non-invasive way some of the above mentioned characteristics leading to promising results in the medical field, in particular to fundamental studies of sample components, characterization and early pathology detection. The filter we present would help to construct more powerful polarimetric techniques based on the Mueller matrix inspection by means of a simple mathematical treatment of $M$.

\begin{figure}[H]
    \centering
    \includegraphics[height=6.5cm]{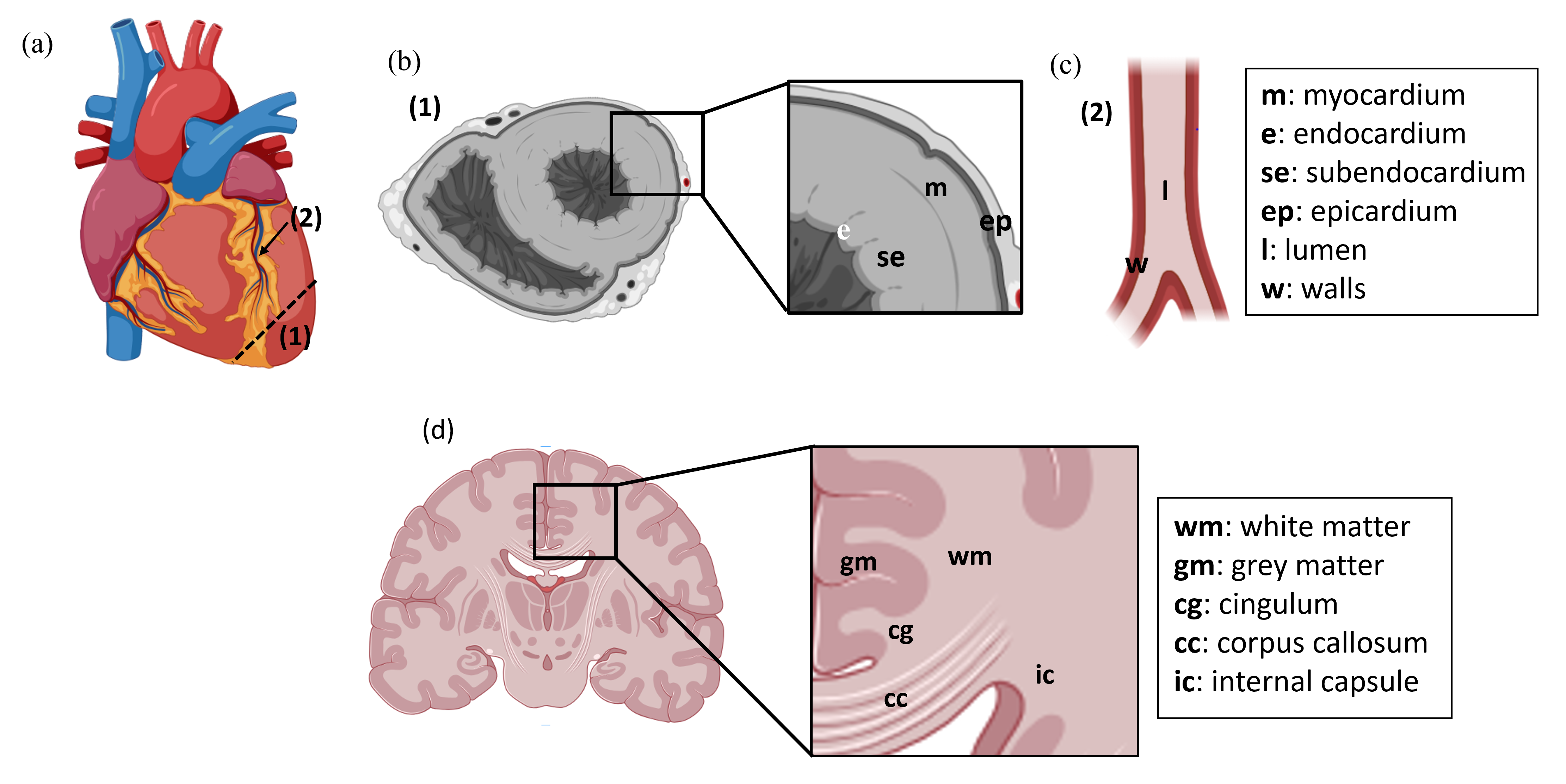}
    \caption{\textbf{(b)}, \textbf{(c)} and \textbf{(d)} correspond to the anatomical context for measurements showed respectively in Fig. \ref{fig:filtromostra_corazon}, Fig. S.6 of the Sup. doc. and Fig. \ref{fig:filtromostra_cerebro}, depicted in humanized schematic drawings. \textbf{(a)} frontal view of a mammalian heart showing the interventricular sulcus containing the interventricular branch of the left coronary artery surrounded by the subepicardial fatty tissue. (1) shows the direction of a section of both ventricles transverse to the heart axis resulting in an image close to the one depicted in (b). In this image the left ventricle appears lined by the endocardium (en) and surrounded by the subendocardium (se), the myocardium (m) and the epicardium (ep). (2) indicates the location of a longitudinal section of the coronary vessel, shown in (c), where (l) and (w) correspond to its lumen and walls respectively. Finally, an idealized drawing of a coronal section (d) of human brain is shown, demonstrating the equivalent areas to the ones analyzed in Fig. \ref{fig:filtromostra_cerebro}.}
    \label{fig:esquema_mostras}
\end{figure}

In the following, we present the results correspondent to a transverse section of  an \textit{ex-vivo} lamb heart sample and a coronal section across the frontal lobe of an \textit{ex-vivo} cattle brain sample. In Fig. \ref{fig:esquema_mostras} we provide a schematic representation of the regions of the samples inspected. We demonstrate how, by means of applying the IDF to the experimentally obtained Mueller matrices, much more information of the samples structures can be obtained.
\subsection{Experimental results for the polarimetric analysis of a heart sample}\label{sec:results_heart}

\begin{figure}[H]
\begin{center}
   \includegraphics[height=10cm]{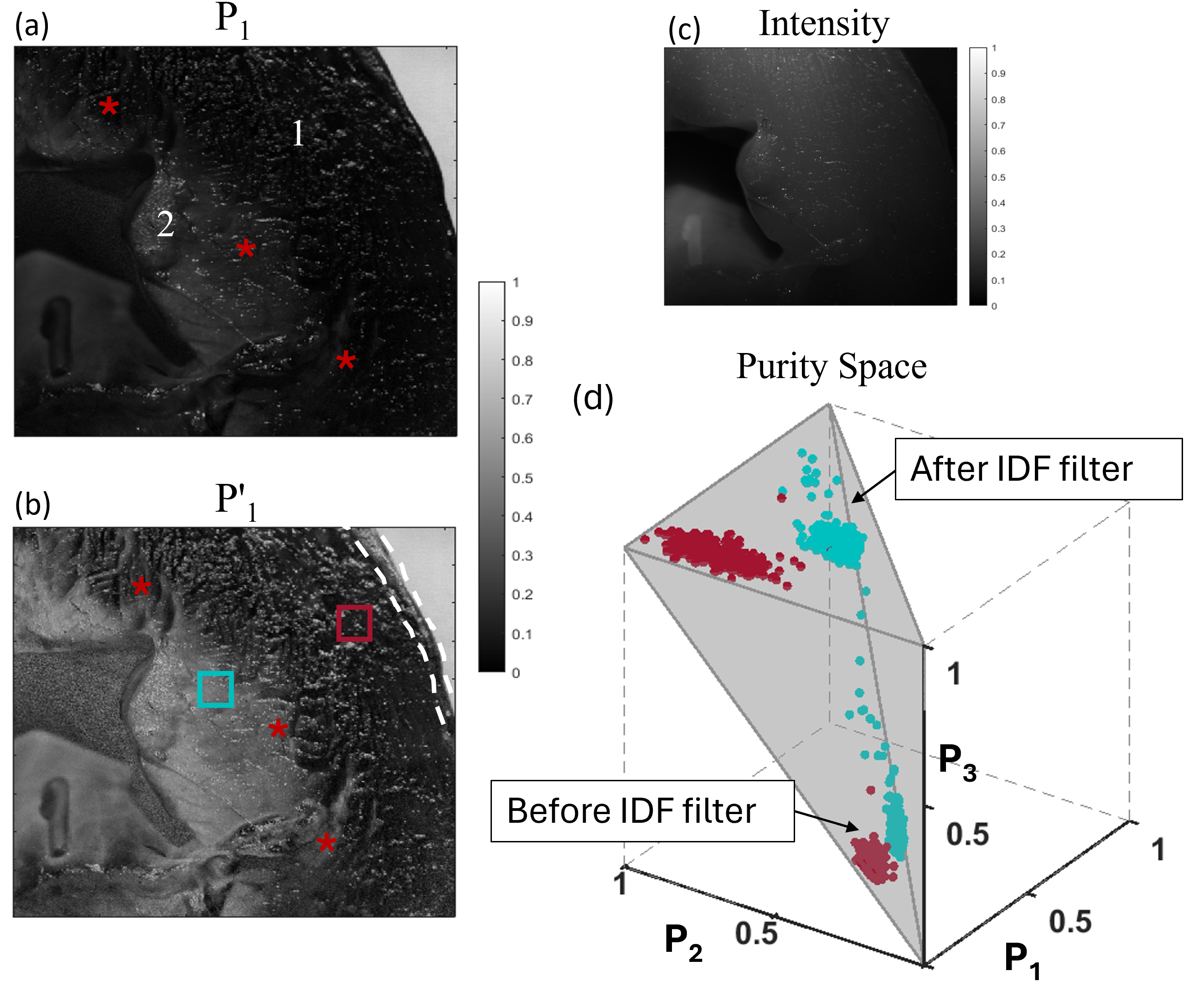}
\caption{Comparison of the polarimetric observable $P_1$ before \textbf{(a)} and after \textbf{(b)} applying the IDF with intensity image \textbf{(c)}. (1) and (2) in (a) indicate myocardial and subendocardial tissue, respectively; white dashed lines in (b) indicate the epicardial tissue and asterisks in (a) and (b) denote some regions where the filter is unveiling information hidden due to isotropic depolarization. (d) corresponds to the representation of the IPP values before and after IDF in the Purity Space; the regions of the sample selected for the Purity Space representation are indicated in (b), red and blue squares correspond to: myocardial and subendocardial tissue, respectively.}
   \label{fig:filtromostra_corazon}
\end{center}  
   \end{figure}

In this section, we present the results of applying the filter to the heart sample. In Fig. \ref{fig:filtromostra_corazon} we compare the standard intensity image (i.e., the non-polarimetric image; Fig. \ref{fig:filtromostra_corazon} (c)) with the images obtained with a particular polarimetric observable, the $P_1$ channel, before (see Fig. \ref{fig:filtromostra_corazon} (a)) and after (see Fig. \ref{fig:filtromostra_corazon} (b)) application of the IDF. We chose $P_1$ for the analysis because, among all polarimetric observables studied, it was the most interesting metric in terms of contrast enhancement and structure unveiling. For the sake of clarity, in Fig. \ref{fig:filtromostra_corazon} (d) we also present a visual interpretation of the effect of the filter on the heart sample data, in terms of the $3D$ Purity Space constructed by means of the IPP.

Figure \ref{fig:filtromostra_corazon} (a) to (c) show the images correspondent to the transverse section of the heart sample. In the conventional  intensity image (unpolarized reflectance), the heart appears as an homogeneous and undifferentiated tissue (see Fig. \ref{fig:filtromostra_corazon} (c)). However, by examining the depolarization response of the sample (through the $P_1$ parameter) we are able to see a spatial dependence, showing different depolarization results for the inner (low grey levels mean low depolarization) and outer region (black means  high depolarization). Interestingly, these regions correspond to different tissue types: (1) is myocardial tissue composed of concentrically arranged fascicles of myocardial muscle whereas (2) corresponds to subendocardial tissue, composed of loose connective tissue and Purkinje fibers. We show how polarimetric analysis reveals structures invisible to conventional techniques. 

Filtering the isotropic depolarization component largely overcomes the performance of both the intensity image and the unfiltered $P_1$ channel. The obtained filtered image is shown in Fig. \ref{fig:filtromostra_corazon} (b), where the contrast between structures (1) and (2) is clearly enhanced compared to the $P_1$ image (in particular, the contrast is enhanced by an amount of $1/P_3$). Furthermore, the application of the IDF leads to the unveiling of new information that was obscured by the isotropic depolarization in the $P_1$ image. In Fig. \ref{fig:filtromostra_corazon} (b), the region corresponding to subendocardial tissue is better defined than before the filter (see Fig. \ref{fig:filtromostra_corazon} (a)). That is, in $P_1'$ we are able to see the border and the two different structures present in this specimen with great contrast. In addition, the filtered image in Fig. \ref{fig:filtromostra_corazon} (b) shows much clearer boundaries and transitions between myocardial and subendocardial tissue, and it is much richer in structural details, as can be seen, for example, in the regions highlighted by red asterisks. Finally, the boundaries between the epicardium (see white dashed lines in Fig. \ref{fig:filtromostra_corazon} (b)) and the myocardial tissue are clearly differentiated, whereas they were almost invisible in both the intensity and $P_1$ images. 

Finally, we note that the IDF is not only useful for enhanced vision, but also for data processing in applications such as tissue classification, as valuable data for training machine learning models. Fig. \ref{fig:filtromostra_corazon} (d), is a graphical representation of the effect of the filter on the values of the IPP represented in their corresponding Purity Space. Complementary information about Purity Spaces can be found in sec. 3 of the supplementary material. Purity Spaces have been shown to be useful in applications related to image-based tissue or structure discrimination \cite{ossikovski2019eigenvalue,rodriguez2021polarimetric}, therefore we find interesting to discuss the effect of the IDF in this framework. To do so, we selected two different regions within the $P_1'$ image, myocardial and subendocardial, these two regions  corresponding to spatial zones of the sample where more information was unveiled after applying the IDF (see blue and red squares in Fig. \ref{fig:filtromostra_corazon} (b)). The values for the pixels within these regions, before and after the filter, are represented in the Purity space with the same color code (see Fig. \ref{fig:filtromostra_corazon} (d)). This representation give us a visual way to understand the beneficial effect of the IDF in terms of data discrimination. Since the $P_3$ parameter controls the height of the tetrahedron describing the Purity Space volume (Fig.  \ref{fig:filtromostra_corazon} (d)), and as the filter effect fixes all values of $P_3$ to 1 (see section 2), we can understand the effect of the filter as the projection of the data clouds, on the plane $P_3=1$ (i.e., all filtered data is re-located to the top surface of the figure). As consequence of this transformation, we can see as the distances between the myocardial and subendocardial heart structures (blue and red datasets) are enlarged after the filter application, this resulting into the subsequent larger discrimination capability between the two regions of the tissue. Moreover, the spread of the points also increases when applying the filter, obtaining richer information of the regions when eliminating the isotropic part of the depolarization, as we shown in Fig.  \ref{fig:filtromostra_corazon}. This increase in the dispersion of the clouds can be quantified, particularly, the dispersion increases by a factor $1/P_3$ (3.92 times larger for this case). 

As a result of this first study case, we demonstrated how by removing the isotropic component of samples depolarization response, by applying the IDF, richer sample information, related to physiological information of tissues, is retrieved. This could be useful for studying myocardial tissue in a more accurate way, helping in the inspection and detection of myocardial tissue modification and leading to the early detection of cardiac diseases such as infarction.

\subsection{Experimental results for the polarimetric analysis of a brain sample} \label{sec:results_brain}
\begin{figure}[H]
\begin{center}
   \includegraphics[height=9.5cm]{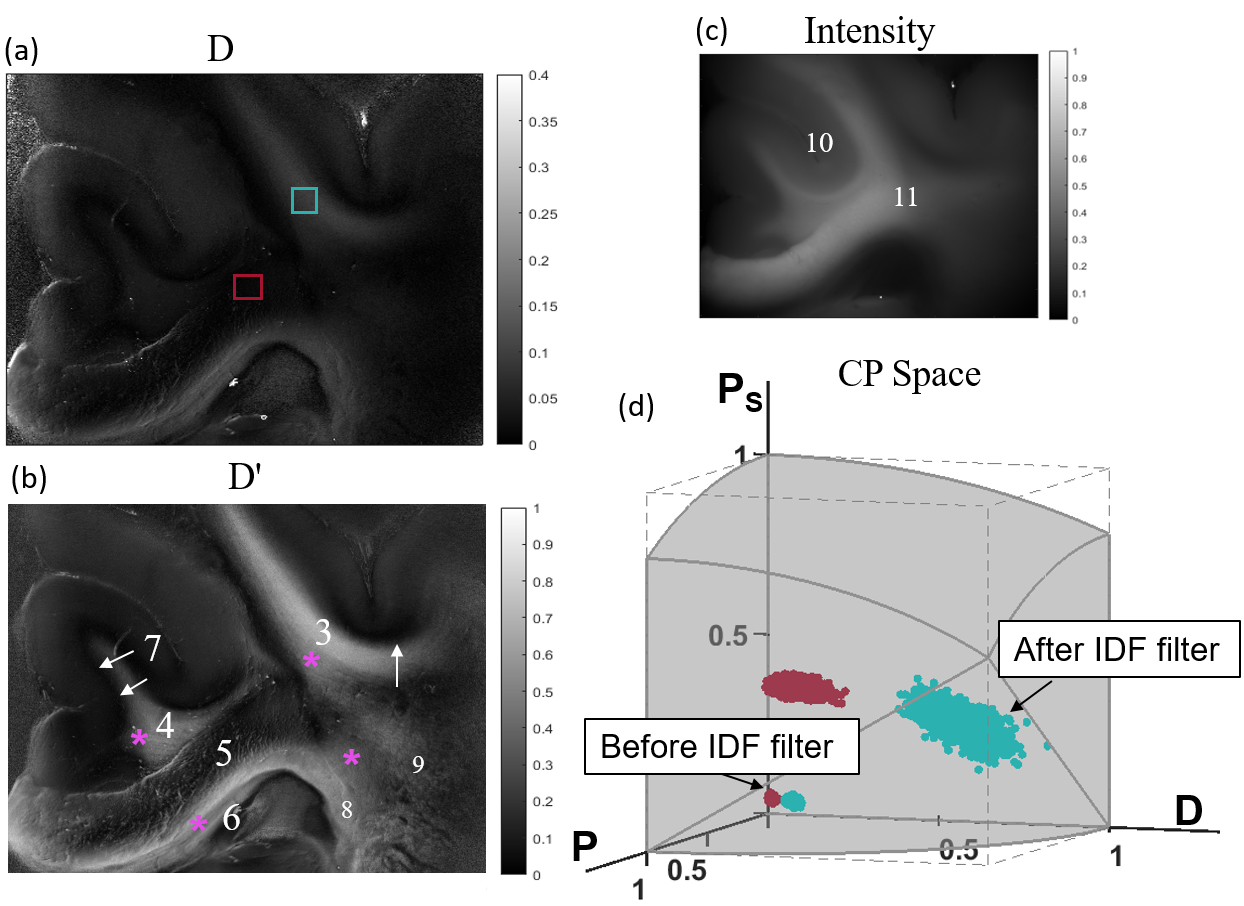}
\caption{ Comparison of the polarimetric observable $D$ before (\textbf{a}) and after (\textbf{b}) applying the IDF with the intensity image (\textbf{c}) in the coronal section across the frontal lobe of the brain. Numbers 3 to 9 in (b) correspond to different white matter tracts of interest. (10) and (11) in (c) denote regions formed by neocortical grey matter and subcortical white matter, respectively.  (d) corresponds to the representation of the CP parameters before and after the filter in the CP Space; the regions of the sample selected for the CP Space representation are indicated in (a), red and blue squares represent different fiber tracts (coronal and rostro-caudal) within WM.}
   \label{fig:filtromostra_cerebro}
\end{center}  
   \end{figure}

We present a second example of the IDF potential by studying its interest for the characterization of a brain sample. Figure \ref{fig:filtromostra_cerebro} shows the comparison between the diattenuation channel before ($D$, Fig. \ref{fig:filtromostra_cerebro} (a)) and after ($D'$, Fig. \ref{fig:filtromostra_cerebro} (b)) being filtered and the non-polarimetric intensity image (Fig. \ref{fig:filtromostra_cerebro} (c)). From all analyzed polarimetric observables images, we selected the diattenuation feature for comparison with standard intensity because it was the metric providing the best results in terms of image enhancement. This result is in agreement with recent studies \cite{menzel2019diattenuation} reporting the interest of diattenuation measurements to reveal very interesting information related to white matter (WM) fiber tracts, this allowing to inspect nerve fiber architecture and distinguish regions with different compositions difficult to detect by means of conventional techniques \cite{agrawal2011josef}. 

 Regarding to the intensity image of the brain section in Fig. \ref{fig:filtromostra_cerebro} (a), we show how the boundaries between neocortical grey matter (10) and subcortical white matter (11) are observed, but no further information of the sample is provided. In turn, when using diattenuation channel $D$ (Fig. \ref{fig:filtromostra_cerebro} (b)), the visualization of brain structures is significantly improved, but the application of the filter on $D$ (Fig. \ref{fig:filtromostra_cerebro} (c)), gives rise to the appearance of significant structures and details, and in particular, the WM tract identification is outstanding, both subcortical and within the corpus callosum (5,6). We also can see clear distinction between tracts following the plane of the section (i.e coronal, or ‘vertical’; 5,7) and tracts, either rostro-caudal (3,4) or medio-lateral (6,8), that do not follow the same plane. Thus, the filtered image allows the identification, in this particularly rich area of the WM, of a wide set of tracts: superior longitudinal fasciculus (3), cingulum (4), medio-dorsal callosal fibers (long-range interhemispheric U-shaped fibers, 5), ventro-lateral (left-right) callosal fibers (6), short-range U-shaped fibers and layer VIb (7, arrows) and ventro-striatal callosal fibers (8). Area termed 9 represents the coalescence between callosal fibers and the internal capsule \cite{mandonnet2018nomenclature}. Pink asterisks in Fig. \ref{fig:filtromostra_cerebro} (b) denote the regions where the filtering effect is revealing hidden information in the polarimetric observable before the filter, detecting information about the fiber directionality masked by isotropic depolarization.
 
The significant brain section visualization improvement achieved by the filtered diattenuation image, $D'$ (Fig. \ref{fig:filtromostra_cerebro} (b)), with respect to the non-filtered image $D$ (Fig. \ref{fig:filtromostra_cerebro} (a)), can be better understood by taking into account the associated range of values. In this sense, as can be see from the corresponding greybars, the diattenuation $D$ values associated to the brain image are restricted between  the range 0-0.4, corresponding this maximum value to the white. That is, the diattenuation response of the sample is low and the range of values that the parameter can reach is quite limited. This implies a lower capability of structure discrimination. Unlike this, when applying the filter, the range variation of the filtered diattenuation $D'$ is largely increased, taking values almost covering the full range (0-1), this leading to the excellent structure unveiling shown in Fig.\ref{fig:filtromostra_cerebro} (b).

Further interpretation of the filter effect on the brain sample data is provided by using a depolarizing Space representation. In particular, Fig. \ref{fig:filtromostra_cerebro} (d) shows the filter effect (data clouds before and after applying the IDF) in terms of the 3D Components of Purity (CP) Space. We used the CP Space since the metric leading to the best results was the diattenuation ($D$) channel, and the CP Space consists of three polarimetric observables, including the diattenuation (in particular, the Diattenuation, $D$, the Polarizance $P$ and the Spherical Purity, $P_s$) \cite{gil2022polarized,van2021unraveling}.

Note that ideal depolarizers (input light is fully depolarized independently of its state of polarization) are located at the point (0,0,0) of the CP Space. As brain data before being filtered presents very low values of the $P_3$ metric, due to their high isotropic depolarizing performance, data clouds are concentrated very close to this point, both for the coronal (in red) and caudal WM fiber tracts (in blue) cases. However, after applying the IDF, isotropic depolarization is removed, so data clouds are displaced far from the fully depolarizing condition, increasing the discriminatory capability between tissues. In particular, when comparing the results before and after filtering represented in Fig. \ref{fig:filtromostra_cerebro} (d) we see outstanding results. The two different regions in the sample (labeled by the red and blue point clouds in the figure) are better discriminated after the application of the filter; not only the separation in the space between the clouds representing different structures is larger but also, the dispersion of the points increases.  Also, in this case the variance of the points increases in a factor 7.57. This situation ensures an excellent performance in terms of tissue discrimination and revealing information of intrinsic properties of the sample. 

Summarizing, we have presented the impactful performance of the IDF for the polarimetric analysis of biological samples. The IDF is particularly useful for the case of soft tissues, where the $P_3$ value is low. However, it can be applied in any kind of sample fulfilling this condition. For the cases presented in this section, we have obtained impressive results, overtaking the response of conventional polarimetric methods. In the heart sample, we shown how the contrast between different structures (myocardial and subendocardial) is highly increased whereas in the brain sample we even reveal structures invisible in the polarimetric images before filtering. These results pave the way to the application of this technique for the early detection of pathologies in heart and brain samples leading to changes in the myocardial structure or the fiber orientation/density, respectively. That could be possible due to the increase of sensitivity in detecting tissue changes that we can get in the filtered polarimetric observables. 

\section{Conclusions}\label{sec:conclusions}
In this work we presented a digital filtering method that takes advantage of different sources of depolarization present in samples. Depolarization can be divided into two types: anisotropic and isotropic. While the former contains information about the physical microscopic constituents of matter, the second one is related with multiple scattering processes in the sample and it can be regarded as a polarimetric white noise. It is important to note that the isotropic depolarization usually hiddens the anisotropic information contained in $M$, so it is interesting to isolate these two contributions. To do this, we inspect the characteristic decomposition of $M$ and see how it can be separated into four Mueller matrices representing different physical systems. In particular, the isotropic depolarization is related to the last term of the decomposition $(1-P_3)M_3$; where $M_3=diag(1,0,0,0)$ is a perfect depolarizer and $1-P_3$ is the weight of the isotropic decomposition. The remaining elements of the decomposition represent the anisotropic properties of the samples. Recall that the proposed isotropic depolarization filter consists in removing this isotropic part of the raw $M$; this leads to a new filtered $M'$ where the anisotropic content is magnified.  

The IDF has been applied to different soft tissues of animal origin, since these structures usually exhibit low $P_3$ values (i.e., they have a high isotropic depolarization response), making them suitable candidates for a successful application of the filter. In particular, we focused the polarimetric analysis on heart and brain samples due to their interest and importance in the medical field, where we demonstrate the outstanding performance of the IDF in terms of structure unveiling and contrast enhancement. In recent years, the potential of polarimetry for biological tissue inspection has already been demonstrated, proving that polarimetric observables provide rich information about sample structure that is not visible with intensity images or conventional medical techniques. At this stage, with the presentation of this IDF, we aim to go one step further by outperforming existing polarimetric techniques. In particular, we show how the filtered polarimetric observables corresponding to the heart samples not only greatly increase the contrast between different tissues (myocardial and subendocardial), but also reveal structures hidden in the conventional polarimetric images in the filtered observables. In the case of the brain, we obtain very interesting results, revealing information about the directionality and identification of fiber tracts in the WM, which were not visible in the unfiltered polarimetric parameters. It is worth noting that we have presented a very simple filtering method for improved visualization of sample features, which can be implemented experimentally using a macroscopic and non-destructive technique. 

Importantly, we want to note that the proposed IDF can be tested with other polarimetric observables than those used in this manuscript. In this sense, we have focused on a reduced set of commonly used observables as representative examples to demonstrate the performance of the filter. However, it could be applied to a large number of already proposed polarimetric observables and/or methods. 
 
The results presented in this work highlight the interest of the filtering method for the study and characterization of biological samples, paving the way for new protocols in biomedical and clinical applications. For example, in the examined samples discussed throughout the manuscript, different structures of interest were revealed. In the case of the heart sample, not only the boundaries and the distinction between myocardial and subenocardial tissue are revealed by IDF, but also the trajectories of the subepicardial coronary vessels and the visualization enhancement of their walls and lumens are obtained. For the brain, we show how IDF can detect and classify tracts with different orientations and resolve the boundaries between grey and white matter. Interestingly, these results can be useful in several medical scenarios. For example, in the study of cardiac pathologies, the inspection and classification of tracts in the white matter can help shed light on the effect of some neuropathies and as a powerful tool for surgeons in the clinical context.

Finally, although we have focused our discussion on the biomedical field, we would like to note that the proposed methods are general and could be useful for a wide range of applications. Note that in all those samples where isotropic depolarization is a significant feature, which is a common situation in several scenarios. For this reason, we invite all researchers dealing with depolarizing samples to test the suitability of the proposed filter.

\appendix
\section* {Data Statement}
Data underlying the results presented in this paper are not publicly available at this time but may be obtained from the authors upon reasonable request.

\section* {Funding sources}

The authors acknowledge the financial support of Ministerio de Ciencia e Innovación and Fondos FEDER (PID2021-126509OB-C21 and PDC2022-133332-C21) and support of Generalitat de Catalunya (2021SGR00138).\\
IM acknowledge the financial support of Ministerio de Ciencia e Innovación and Fondos FEDER (PID2021-126509OB-C22).
\section*{Author contributions}

MCC conducted the experiments, developed the concept, and analyzed experimental data and wrote the results. EGA helped in experimental part and the writing. AB analyzed experimental data. IMG helped in the edition and writing. JJG supervised the the theoretic part. EGC and RO participated in the discussions of the results and revision of the article. IM helped in the discussion of the results and writing and editing process. IE, JC and AL supervised and helped in the idea development, analysis and writing process. JC provided the financial support. 
\section*{Competing interests}
The authors declare no conflicts of interest.

\section*{Supplemental document}
See Supplement 1 for supporting content. 

\bibliographystyle{elsarticle-num} 
\bibliography{report}

\newpage

\section*{\LARGE{\textbf{Supplementary Document}}}

\section{Covariance matrix (H)}\label{sec:covmatrix}

The covariance matrix $H$ is a Hermitian semi-definite matrix that arises from a transformation of the Mueller matrix ($M$)\cite{gil2022polarized}:
\vspace{-1mm}
\begin{equation}\tag{S.1}
    H = \frac{1}{4} \sum_{i,j} m_{ij} (\sigma_i \otimes \sigma_j),
\end{equation}

\hspace{-6mm}where $m_{ij}$ represent the $M$ elements, $\sigma$ are the Pauli matrices, and $\otimes$ is the Kronecker product. This transformation of $M$ to $H$ is convenient since $H$ is a Hermitian matrix, and thus diagonalizable, whereas $M$ is not necessarily diagonalizable. The Indices of Polarimetric Purity (IPP) arise from lineal combinations of $H$ eigenvalues. Interestingly, the eigenvalues of $H$ connect with the enpolarizing (polarizing and depolarizing) properties of samples \cite{canabal2024connecting}.

To fulfill the condition of a physically realizable $M$, H eigenvalues must satisfy the Cloude’s criterion \cite{cloude1986group}:
\vspace{-1mm}
\begin{equation}\tag{S.2}
    \lambda_4 \leq \lambda_3 \leq \lambda_2 \leq \lambda_1 \leq 0,
    \label{eq:lambdas}
\end{equation}

\begin{equation}\tag{S.3}
    \sum_{i=1}^4\lambda_i =1.
    \label{eq:sum_lambdas}
\end{equation}

The combination of Eqs. (\ref{eq:lambdas}) and (\ref{eq:sum_lambdas}) allow to eliminate one eigenvalue, resulting in:

\begin{equation}\tag{S.4}
    0 \leq \lambda_4 \leq \lambda_3 \leq \lambda_2 \leq 1 - \lambda_4 - \lambda_3 - \lambda_2.
\end{equation}

\section{Effect of the IDF on polarimetric observables}\label{sec:effectIDF}

In this section, we discuss the effect of the isotropic depolarization filter (IDF) in terms of image contrast when applied to certain polarimetric observables (non depolarizing \ref{subsec:nondepchannels} and depolarizing \ref{subsec:depchannels}).

\subsection{Non-depolarizing channels}\label{subsec:nondepchannels}

The dichroic properties of a sample can be represented by the diattenuation ($D$) and polarizance ($P$) metrics. These metrics can be directly extracted from $M$. In particular, by writing the Mueller matrix in its block form \cite{gil2022polarized}, we obtain the following expression: 

\begin{equation}\tag{S.5}
    M = m_{00}\begin{pmatrix}
        1 & D^T \\
        P & m
    \end{pmatrix},
\end{equation}

\begin{equation}\tag{S.6}
    D = \frac{\sqrt{m_{01}^2 + m_{02}^2 + m_{03}^2}}{m_{00}}, \quad P = \frac{\sqrt{m_{10}^2 + m_{20}^2 + m_{30}^2}}{m_{00}},
\end{equation}

\hspace{-6mm} where $m_{ij}$ (i,j=0,…3) are elements of $M$ and $m$ denotes the 3x3 submatrix of $M$. To obtain the effect of the filter on these parameters, we need to know the response of the $M$ elements to the filter. As we show in the main manuscript, the only element of $M$ affected by the filter is $m_{00}$ (see Eq. (6) of the main text). Therefore, the expressions for $D$ and $P$ after the filter, i.e. $D’$ and $P’$, are

\begin{equation}\tag{S.7}
    D' = \frac{\sqrt{m_{01}^2 + m_{02}^2 + m_{03}^2}}{P_3 m_{00}}, \quad P' = \frac{\sqrt{m_{10}^2 + m_{20}^2 + m_{30}^2}}{P_3 m_{00}}.
    \label{eq:pd}
\end{equation}

\subsection{Depolarizing channels}\label{subsec:depchannels}

In this text, we present different depolarizing metrics: the IPP and the indices $Ps$ and $P_{\Delta}$. The spherical purity index $P_s$ is calculated as follows\cite{gil2022polarized}:

\begin{equation}\tag{S.8}
    P_s = \frac{\|m\|_2}{\sqrt{3}},
\end{equation}

\hspace{-6mm}where $\|m\|_2$ is the Frobenius norm of $M$. In the case of the depolarization index, it can also be obtained in terms of the $M$ elements. However, for this study we find more interesting to define this index by its relationship with different polarimetric parameters. The $P_{\Delta}$ parameter can be calculated both by means of the IPP and the so-called Components of Purity (CP), comprised by the dichroic parameters previously described $D$ and $P$ and $Ps$:

\begin{equation}\tag{S.9}
    P_\Delta = \frac{\sqrt{D^2 + P^2 + P_s^2}}{3}=\frac{\sqrt{6P_1^2 + 2P_2^2 + P_3^2}}{3}, 0 \leq P_{\Delta} \leq 1.
    \label{eq:pdelta}
\end{equation}

For the case of $P_s$, the effect of the filter affects the value of the Frobenius norm in the same way as that shown for the dichroic parameters ($\|m'\|_2=\|m\|_2/P_3$): 

\begin{equation}\tag{S.10}
    P_s' = \frac{\|m'\|_2}{P_3\sqrt{3}}=\frac{P_s}{P_3}.
    \label{eq:psfiltrada}
\end{equation}

Finally, the effect of the filter on the $P_{\Delta}$ index can be directly calculated by replacing the metrics in Eq. \ref{eq:pdelta} by its filtered version:

\begin{equation}\tag{S.11}
P_\Delta' = \frac{\sqrt{D^2 + P^2 + P_s^2}}{3}=\frac{\sqrt{6P_1^2 + 2P_2^2 + P_3^2}}{3}=\frac{P_s}{P_3}.
\end{equation}

\section{Depolarization spaces}\label{sec:dep_spaces}

\subsection{Purity Space}
The Purity Space is the volume generated by the IPP triplet, the constrains among the three variables ($0 \leq P_1\leq P_2\leq P_3\leq 1$) generate a tetrahedron containing all the physically realizable depolarizers \cite{ossikovski2019eigenvalue,gil2022polarized}. The effect of the filter in contrast enhancement and visualization for the IPP parameters is clear when inspecting this space (from a given value of $P_3$; see Fig. S1. According to the above stated inequality, the index $P_3$ governs the maximum value of $P_2$ and $P_1$. By applying the filter, $P_3$ value is fixed to 1 (maximum value), and thus, $P_1$ and $P_2$ indices can have values between 0 and 1 (not only from 0 and a smaller limit set by the $P_3$ value). Note that different values for $P_3$ lead to different surfaces in the Purity Space volume (see Fig. S1). Thus, as larger the $P_3$ value, as larger the surface area (see the surfaces represented by different colors in Fig. S1, corresponding to different values of $P_3$). 
Fig. S1 provides a representation of the Purity Space, where the $P_3$ metric corresponds to the z axis (height of the tetrahedron) and $P_1$ and $P_2$ correspond to the x and y axis. For a fixed value of $P_3$ (that is, for a particular z-plane in the tetrahedron) the surface delimited by $P_1$ and $P_2$ increases as $P_3$ increases (see different colors codifying different $P_3$ values in Fig. S1). Therefore, by applying the IDF to a sample, the polarimetric data associated with such a sample is represented in a larger surface of the tetrahedron. Under this scenario, as smaller are the values of $P_3$ associated with a sample, as larger will be the potential of the IDF and, larger information of diattenuation and retardance encoded in $P_1$ and $P_2$ metrics could be extracted. What is more, the potential of the IDF to largely separate between different depolarizers in depolarizing spaces implies an increase on the visualization and discriminatory capabilities of $P_1$ and $P_2$ observables, this being the major goal and contribution of the filter. 
\vspace{-2mm}
\begin{figure}[H]
    \centering
    \includegraphics[width=0.65\textwidth]{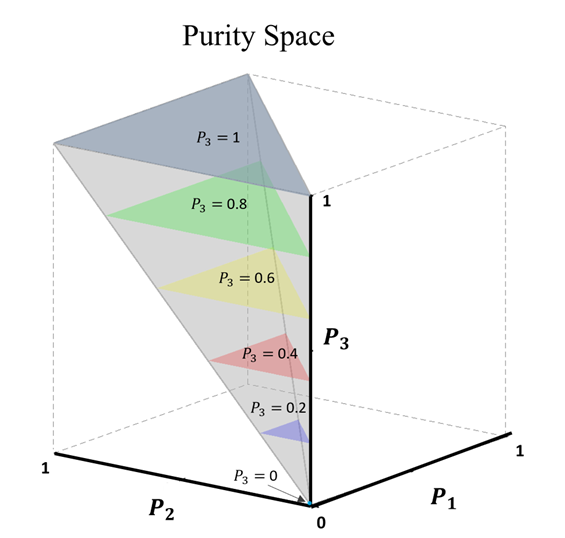}
    \caption*{Figure S1: Representation of the IPP space, where all the physically realizable depolarizers are contained. Color surfaces represent the possible values that the indices $P_1$ and $P_2$ can achieve for a particular constant value of $P_3$ ($P_3 = 1, 0.8, 0.6, 0.4, 0.2, 0$ corresponding to grey, green, yellow, red, violet, and blue colors, respectively).}
    \label{fig:$P_3$_space}
\end{figure}

\subsection{Components of Purity Space}

In analogy to Purity Space, we study the effect on the space conformed by the Components of Purity or Purity Figure, which contains complementary information about the depolarization properties of a sample \cite{van2021unraveling}. The CP space consist of the diattenuation ($D$), polarizance ($P$) and degree of spherical purity ($Ps$) metrics (the later connected with depolarizing origin not related with dichroism). The CP Space represents a volume where all the possible depolarizers are contained. The constraint between the three components giving rise to this space is the following \cite{gil2022polarized}:

\begin{equation}\tag{S.12}
     P^2 + D^2 \leq 1+ P_s^2.
\end{equation}

In the CP Space (see Fig. S2), the maximum achievable surface is governed by the $P_{\Delta}$ value, that is, as larger $P_{\Delta}$, as larger the correspondent surface in the volume, where a larger surface corresponds to a better discriminatory capability. In this scheme, the less depolarization in a sample ($P_{\Delta}$ = 1 for a non-depolarizing media and $P_{\Delta}$ = 0 for a totally depolarizing media), the larger discriminatory capability of the surface (see Fig. S2).

The effect of the IDF in the triplet conforming this space is the same as for the IPP, where the value increases by $1/P_3(x,y)$ (see Eqs. 
(\ref{eq:pd}) and (\ref{eq:psfiltrada}). Whereas in Purity Space the maximum surface in the volume (for a fixed $P_3$) increase with the $P_3$ value, for the CP space this increase is related to the value of $P_{\Delta}$ \cite{van2021unraveling}. The increase of $P_{\Delta}$ after the filter (see Eq. (\ref{eq:pdelta})) has the direct effect in the CP Space of increasing the possible surface that the parameters can occupy (see Fig. S2). Therefore, we demonstrate how the IDF can be also very convenient for these metrics, the contrast enhancement and visualization of different structures inside a sample can be significantly improved and hidden structures due to the presence of a high amount of isotropic depolarization can be revealed. 
\vspace{-2mm}
\begin{figure}[H]
    \centering
    \includegraphics[width=0.65\textwidth]{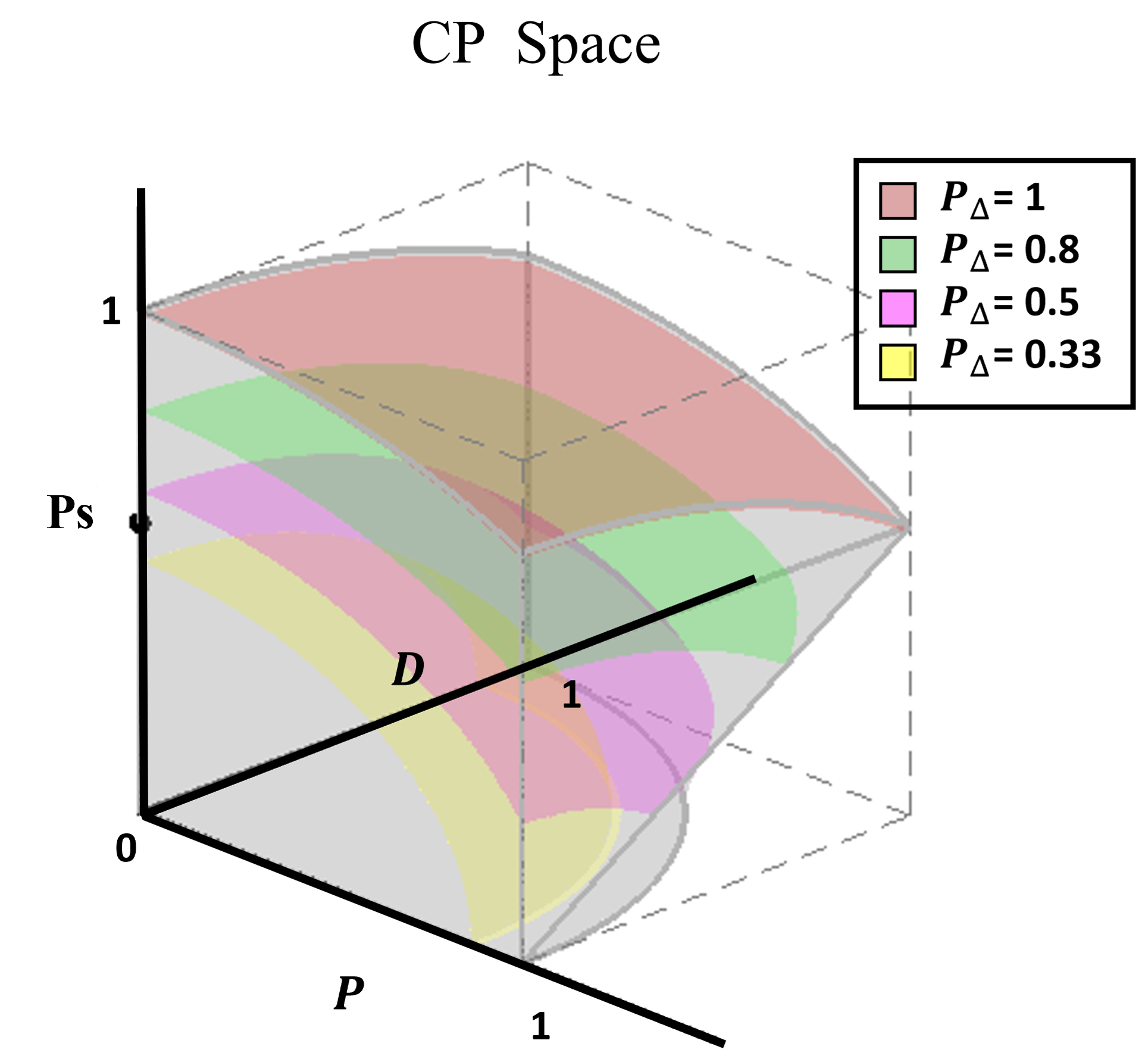}
    \caption*{Figure S2: Representation of the Components of Purity space. The color surfaces represent the regions associated to the different $P_{\Delta}$ values ($P_{\Delta} = 1, 0.8 ,0.5, 0.33$, corresponding to red, green, pink and yellow colors, respectively).}
    \label{fig:cp_space}
\end{figure}
\vspace{-1mm}
We want to highlight by applying the proposed IDF on samples on the metrics associated to the IPP and CP spaces, in both cases, as larger the isotropic content of samples, as larger the potential of the IDF in terms of imaging contrast. Importantly, isotropic depolarization is a common response of biological tissues, as it is the case of study in our work, but the filter is general and could be useful for any sample showing isotropic depolarizing performance.

\section{Experimental setup: Mueller matrix polarimeter}\label{sec:exp_setup}

To obtain the experimental $M$ of the samples inspected in this text we use a complete image Mueller polarimeter. These polarimeters are comprised by two main parts: the polarization state generator (PSG) and the polarization state analyzer (PSA), which allow to generate and analyze, respectively, any state of polarization. The PSG is comprised of a linear polarized oriented at 0$^{\circ}$  with respect to the laboratory vertical and two parallel aligned liquid crystal (PA-LC) retarders oriented at 45$^{\circ}$ and 0$^{\circ}$, respectively. In the case of the PSA, the elements conforming this system are the same as in the PSG but located in inverse order. Moreover, to obtain M images a CCD camera is also located after the PSA system, capturing the intensity of the sample corresponding to each one of the camera pixels. With the elements comprising the PSG and the PSA we can generate and analyze any state of fully polarized light. In addition, the system is provided with a LED source located before the PSG for illumination of the sample; this source can work at three illumination wavelengths in the visible range (625 nm, 530 nm and 470 nm) allowing to inspect different characteristics of the samples. Both arms comprising the polarimeters are located in rotation stages, therefore the angle configuration can be changed. In this work, we set the reflection configuration for measuring the light reflected by the samples: the sample is illuminated with the PSG located at 34$^{\circ}$ with respect to the laboratory horizontal and the PSA is located at 0$^{\circ}$ with respect to the laboratory vertical to avoid direct reflections. A visual representation of the setup is shown in Fig. S3.

\begin{figure}[H]
    \centering
    \includegraphics[width=0.9\textwidth]{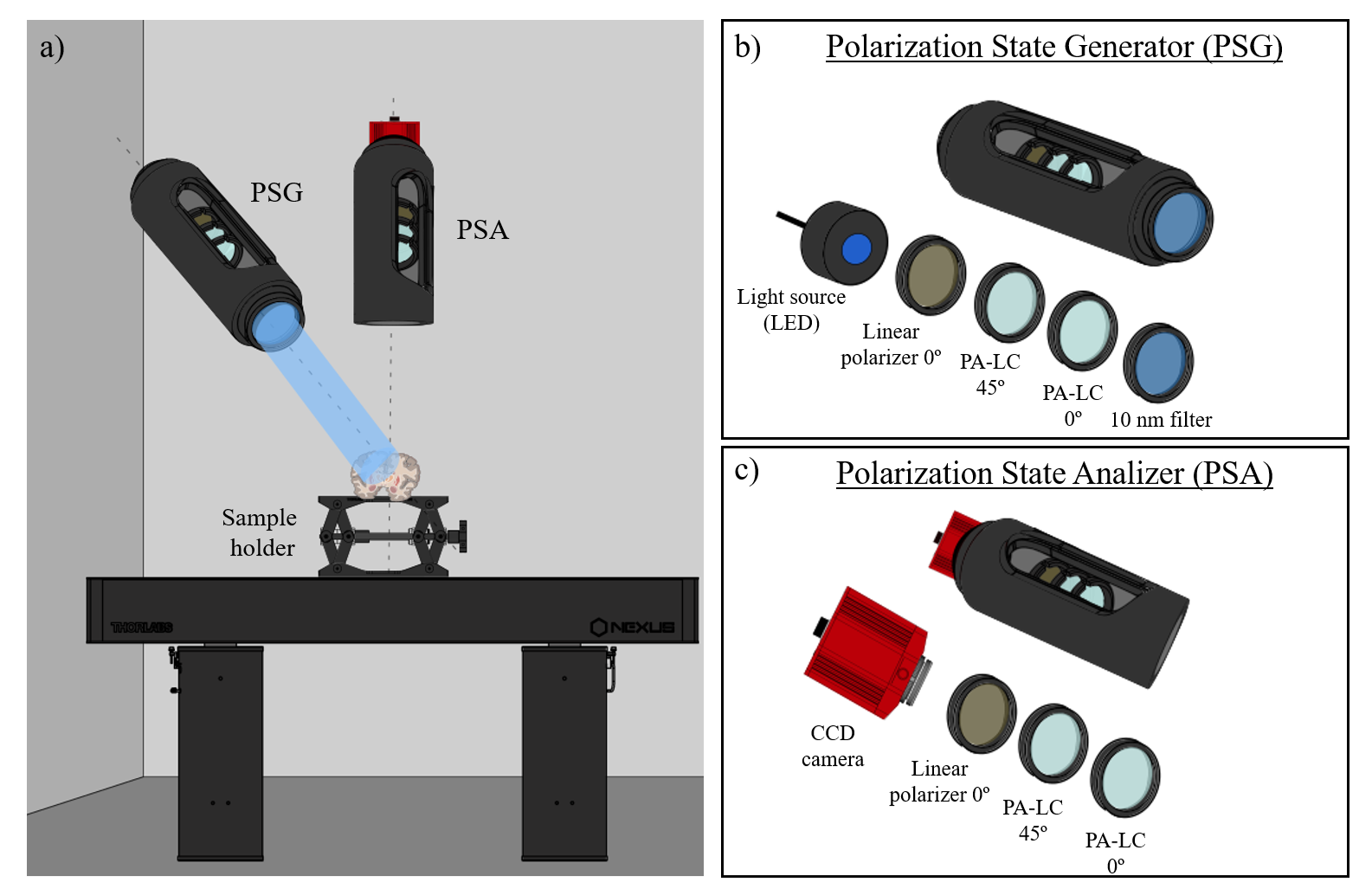}
    \caption*{Figure S3: (a) Representation of the complete image Mueller polarimeter at scattering configuration. Representation of the polarization state generator (PSG; b) and polarization state analyzer (PSA; c) optical components arrangement.}
    \label{fig:experimental_setup}
\end{figure}
We also provide detailed information about the components comprising the polarimeter: the illumination is provided by a Thorlabs LED source (LED4D211, operated by DC4104 drivers distributed by Thorlabs) complemented with a 10 nm dielectric bandwidth filters distributed by Thorlabs: FB530-10 and FB470-10 for green and blue wavelengths, respectively. The linear polarizer located in the PSG is a Glam–Thompson prism-based CASIX whereas the one placed in the PSA is a dichroic sheet polarizer distributed by Meadowlark Optics. The four PA-LC retarders are variable retarders with temperature control (LVR-200-400-700-1LTSC distributed by Meadowlark Optics). Finally, imaging is performed by means of a 35 mm focal length Edmund Optics TECHSPEC high resolution objective followed by an Allied Vision manta G-504B CCD camera, with 5 Megapixel GigE Vision and Sony ICX655 CCD sensor, 2452(H)×2056(V)2452(H)×2056(V) resolution, and cell size of $3.45\mu m \times 3.45\mu m \times 3.45\mu m \times 3.45\mu m$, so a spatial resolution of $22\mu m$ is achieved. 
\newpage
\section{Experimental Results}\label{exp_results}

In this section we provide further discussion related to some experimental studies we conducted to validate the potential of the filter.

\subsection{Mean IPP values for various animal samples}

\begin{table}[H]

\label*{Table S1: IPP values for different animal samples before applying the IDF.}
\centering
\begin{tabular}{|ccccc|}
\hline
\rowcolor[HTML]{D5DCE4} 
Sample                                                                                           & Wavelength (nm)             & $P_1$     & $P_2$     & $P_3$     \\ \hline
Skin (pork)                                                                                      & \cellcolor[HTML]{FBE4D5}625 & 0.0658 & 0.0976 & 0.1313 \\ \hline
Brain WM (cow)                                                                                   & \cellcolor[HTML]{DEEAF6}470 & 0.0268 & 0.0413 & 0.1249 \\ \hline
Brain GM (cow)                                                                                   & \cellcolor[HTML]{DEEAF6}470 & 0.1552 & 0.1943 & 0.4201 \\ \hline
                                                                                                 & \cellcolor[HTML]{FBE4D5}625 & 0.0718 & 0.2019 & 0.2491 \\
                                                                                                 & \cellcolor[HTML]{E2EFD9}530 & 0.1307 & 0.2037 & 0.3257 \\
\multirow{-3}{*}{Epiglottis (cattle)}                                                            & \cellcolor[HTML]{D9E2F3}470 & 0.1541 & 0.2109 & 0.3468 \\ \hline
                                                                                                 & \cellcolor[HTML]{FBE4D5}625 & 0.0886 & 0.1836 & 0.2451 \\
                                                                                                 & \cellcolor[HTML]{E2EFD9}530 & 0.1887 & 0.3463 & 0.4550 \\
\multirow{-3}{*}{\begin{tabular}[c]{@{}c@{}}Tong (cattle)\\      muscle exterior\end{tabular}} & \cellcolor[HTML]{DEEAF6}470 & 0.1584 & 0.2858 & 0.4202 \\ \hline
                                                                                                 & \cellcolor[HTML]{FBE4D5}625 & 0.0437 & 0.1292 & 0.1971 \\
                                                                                                 & \cellcolor[HTML]{E2EFD9}530 & 0.0760 & 0.1290 & 0.2222 \\
\multirow{-3}{*}{\begin{tabular}[c]{@{}c@{}}Tong (cattle)\\     muscle interior\end{tabular}} & \cellcolor[HTML]{D9E2F3}470 & 0.0623 & 0.1134 & 0.2434 \\ \hline
Heart (cattle) myocardium                                                                        & \cellcolor[HTML]{FBE4D5}625 & 0.1321 & 0.0753 & 0.0320 \\ \hline
\end{tabular}
\end{table}

\subsection{Application of the IDF in biological samples}

In this section we present the results of applying the IDF to different biological samples. In particular, two sections of \textit{ex-vivo} lamb heart (samples 1 and 3) and two sections of \textit{ex-vivo} cattle brain (samples 2 and 4). Samples 1 and 2 were described in the main text; sample 3 corresponds to a zenithal, external, view of the interventricular region of a cattle heart and sample 4 to a coronal section of cow occipital lobe. In the following, we present a set of polarimetric observables before and after filtering to show the potential of the method in different metrics for samples 1 to 4. 
Fig. S4 to Fig. S7 show the results of the IDF in the four mentioned samples. For all the images, a comparison between a set of representative polarimetric observables before and after the application of the filter is provided. The polarimetric observables presented in each case correspond to images leading to the best results in terms of visualization enhancement and structure unveiling. 

\begin{figure}[H]
    \centering
    \includegraphics[width=0.8\textwidth]{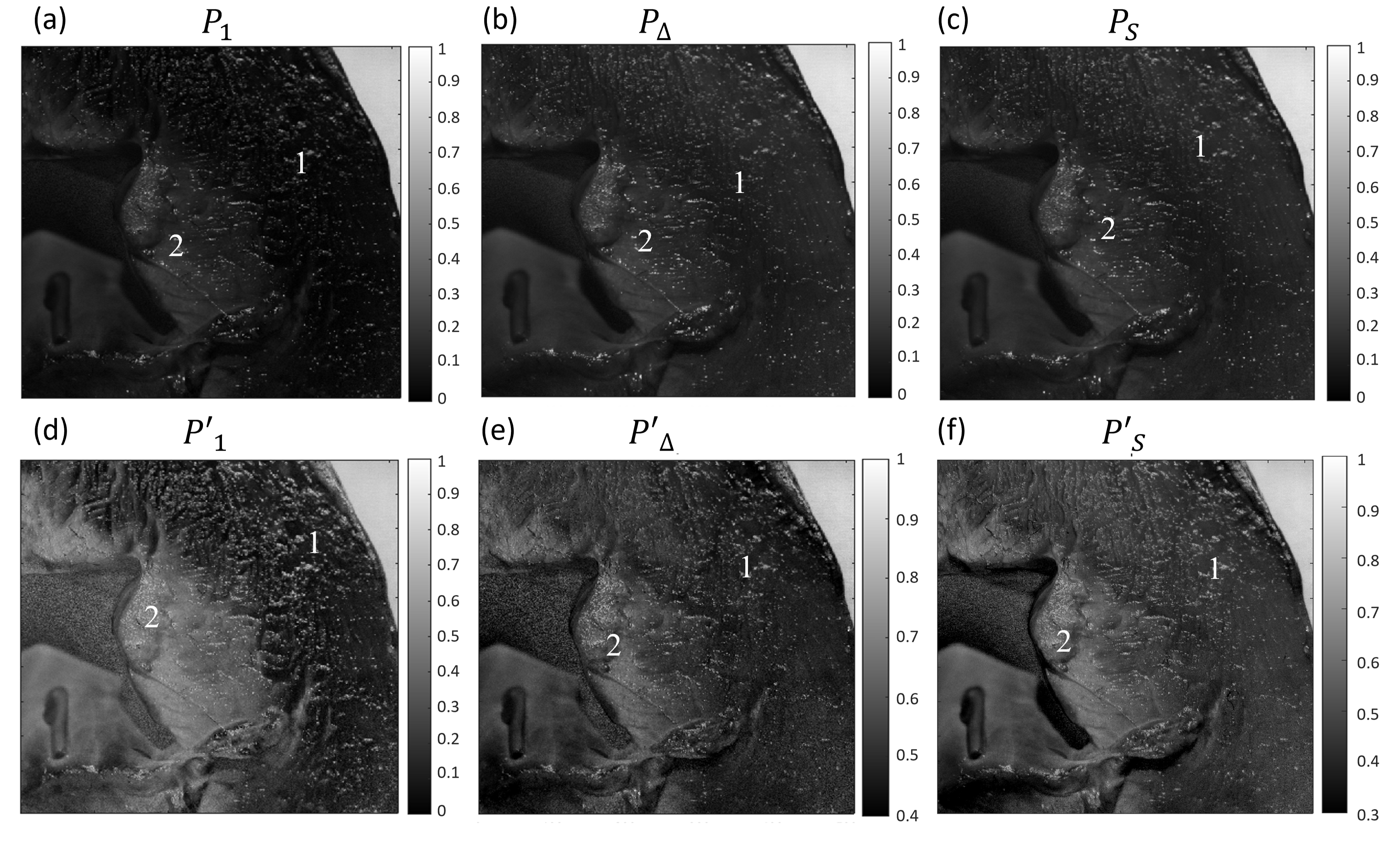}
    \caption*{Figure S4: Sample 1: transverse section of cattle left ventricle. Comparison between polarimetric observables before ((a)-(c)) and after ((d)-(f)) applying the IDF arising from $P_1$, $P_{\Delta}$ and Ps, respectively. All pairs of images show enhanced contrast between myocardium (1) and subendocardium (2) after filtering.}
    \label{fig:corazon1}
\end{figure}

\begin{figure}[H]
    \centering
    \includegraphics[width=0.9\textwidth]{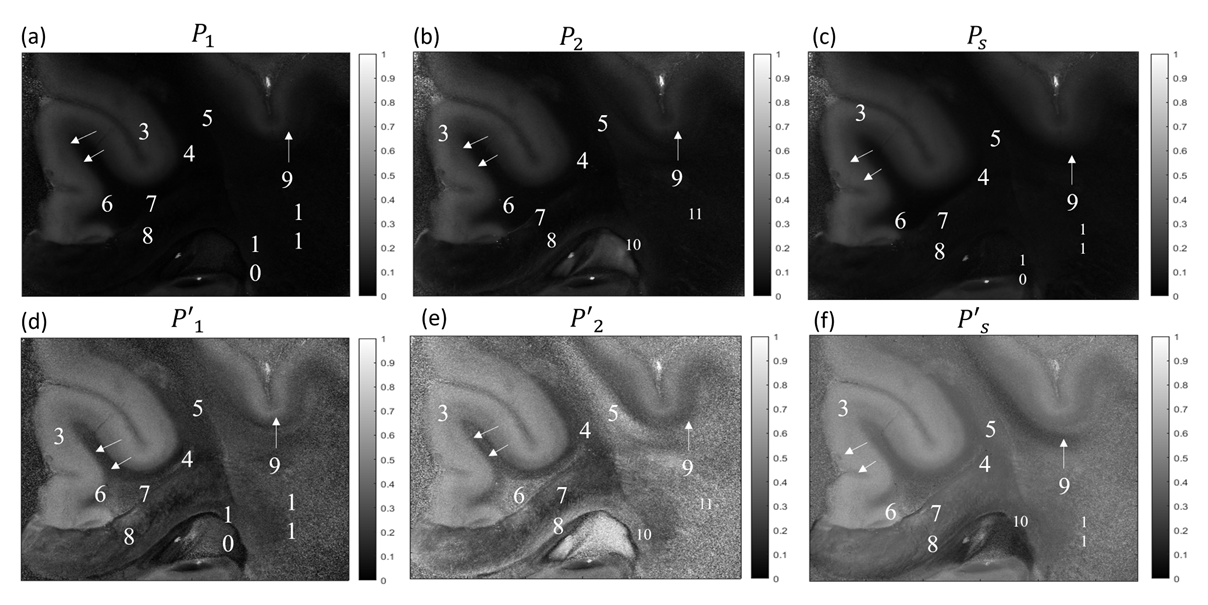}
    \caption*{Figure S5: Sample 2: coronal section of cow frontal lobe. Comparison between polarimetric observables before ((a)-(c)) and after ((d)-(f)) being filtered for $P_1$, $P_2$ and $Ps$, respectively. All pairs of images show maintained contrast between grey matter (GM) and white matter (WM) (3 for GM ,4 for WM), while filtered $P_2$ allows discrimination between individual fascicles within subcortical and callosal WM: superior longitudinal fasciculus (5), cingulum (6), medio-dorsal callosal fibers (long-range interhemisferal U-shaped fibers, 7), ventro-lateral (left-right) callosal fibers (8), short-range U-shaped fibers and 6 layer (9, arrows) and ventro-striatal callosal fibers (10). Area termed 11 represents the coalescence between callosal fibers and the internal capsule.}
    \label{fig:cerebro1}
\end{figure}

\begin{figure}[H]
    \centering
    \includegraphics[width=0.9\textwidth]{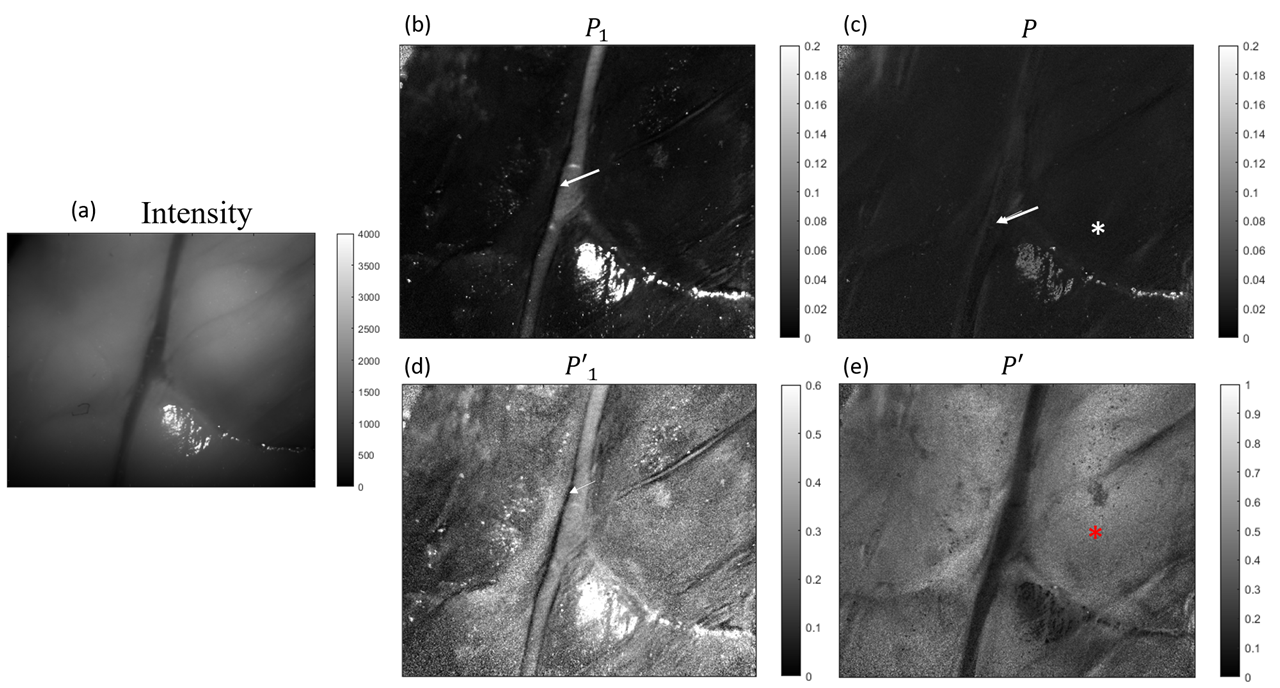}
    \caption*{Figure S6: Sample 3: zenithal, external, view of the interventricular region of a cattle heart. Comparison between intensity (a) and polarimetric observables before ((b)-(c)) and after ((d)-(e)) being filtered for $P_1$, $P_2$ and $Ps$ respectively. Filtered $P_1$ allows the identification of the endotelial-muscular lining of a subepicardial vessel (arrow in (d)), while filtered P reveals the distribution of the subepicardial fatty tissue (red asterisk in ‘e’).}
    \label{fig:corazon2}
\end{figure}

\begin{figure}[H]
    \centering
    \includegraphics[width=1\textwidth]{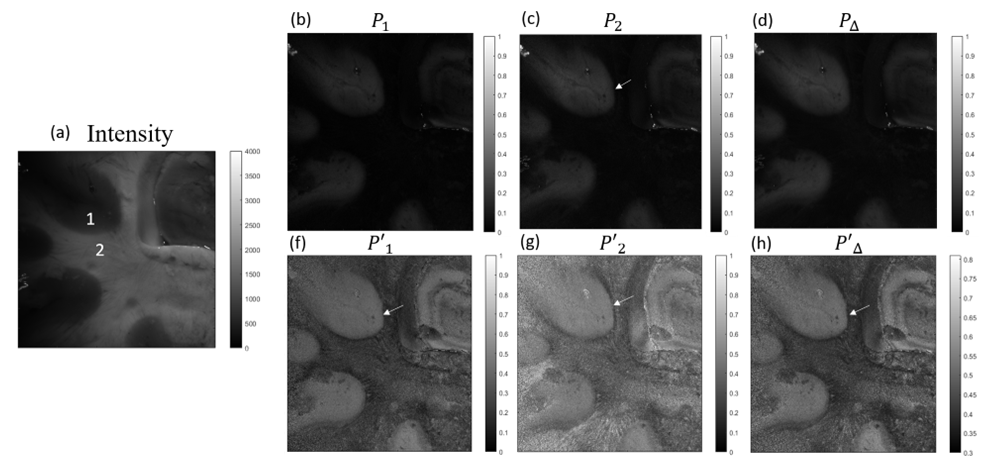}
    \caption*{Figure S7: Sample 4: coronal section of cattle occipital lobe. Comparison between intensity (a) and polarimetric observables before ((a)-(d)) and after ((f)-(h)) being filtered for $P_1$, $P_2$ and $P_{\Delta}$, respectively. Although occipital tracts are less defined and varied in directionality than frontal ones (see Fig. 3) all pairs of images show maintained contrast between GM (1) and WM (2) as well as identification of short-range U-shaped fibers and VIb layer (arrows) that are not readily identifiable in unfiltered images.}
    \label{fig:cerebro2}
\end{figure}

\end{document}